\documentstyle[aas2pp4,epsf]{article}

\lefthead{}
\righthead{}

\def\Msun{M_{\sun}}
\def\cm3{{\rm ~cm}^{-3}}
\def\ltsima{$\; \buildrel < \over \sim \;$}
\def\ltsim{\lower.5ex\hbox{\ltsima}}

\begin{document}
\title{EVOLUTION OF BERYLLIUM AND BORON IN THE INHOMOGENEOUS EARLY GALAXY}

\author{Takeru Ken Suzuki$^{1,2}$, Yuzuru Yoshii$^{3,4}$, and
Toshitaka Kajino$^{2,5,1}$}

\altaffiltext{1}{Department of Astronomy, School of Science,
University of Tokyo, Bunkyo-ku, Tokyo, 113-0033 Japan}
\altaffiltext{2}{Theoretical Astrophysics Division, National
Astronomical Observatory, Mitaka, Tokyo, 181-8588 Japan;
stakeru@yso.mtk.nao.ac.jp, kajino@th.nao.ac.jp}
\altaffiltext{3}{Institute of Astronomy, School of Science, 
University of Tokyo, Mitaka, Tokyo, 181-8588 Japan; 
yoshii@ioa.s.u-tokyo.ac.jp}
\altaffiltext{4}{Research Center for the Early Universe, School 
of Science, University of Tokyo, \\Bunkyo-ku, Tokyo, 113-0033 Japan}
\altaffiltext{5}{Department of Astronomical Science, Graduate 
University for Advanced Studies, Mitaka, Tokyo, 181-8588 Japan}

\begin{abstract}

A model of supernova-driven chemical evolution of the Galactic halo, 
recently proposed by Tsujimoto, Shigeyama, \& Yoshii (1999, ApJL, 519,
63), is extended in order to investigate the evolution of light elements 
such as Be and B (BeB), which are produced mainly through spallative reactions
with Galactic cosmic rays (GCRs).  In this model each supernova (SN) sweeps up
the surrounding interstellar gas into a dense shell and directly enriches it
with ejecta which consist of heavy elements produced in each Type II supernova
with different progenitor masses.  We propose a two-component source for GCRs
such that {\it both} interstellar gas and fresh SN ejecta engulfed in the shell
are accelerated by the shock wave.  The released GCRs travel much faster than
the expansion of the shell, and thus produce the BeB elements far outside the
shell which will be incorporated in subsequent formation of shells arising from
later SNe.  As a result, stars formed from coeval shells are predicted to
show a large scatter in the abundance of heavy elements, while exhibiting BeB
abundances similar to that in the gas, with no appreciable scatter.  This
indicates that, contrary to heavy elements, stellar BeB abundances might be
used as a good age indicator in the inhomogeneous Galactic halo.

The production of BeB at early epochs is dominated by the primary process
through spallation of heavy GCRs, though being a minor component in the bulk of
the GCR composition at present.  We have calculated the frequency distribution
of long-lived stars in the log(BeB/H)-[Fe/H] plane and find that the contour
of constant frequency covering a range of $-3<$[Fe/H]$<-1$ in this plane is
consistent with the observed linear trend between BeB and Fe.  We
show from our calculations that there is an intriguing possibility of
distinguishing between standard and non-standard Big-Bang nucleosynthesis
models if BeB abundances in several hundred halo stars are observed in
the future.

\end{abstract}


\section{INTRODUCTION}

The study of elemental abundance patterns in very metal-poor halo stars has
suggested that stars with $-4<$[Fe/H]$<-2.5$ were born from individual
supernova remnants (SNR), reflecting a different nucleosynthesis yield for each
Type II supernova (SN II) with different progenitor mass (Shigeyama \&
Tsujimoto 1998, hereafter ST98; Ishimaru \& Wanajo 1999).  This indicates that
previous models of chemical evolution in which stars are assumed to form from
well-mixed, uniform gas cannot apply to the Galactic halo; as a result
a unique age versus metallicity relation is not expected to hold for
stars with [Fe/H]$<-2.5$.  Tsujimoto, Shigeyama, \& Yoshii (1999, hereafter
TSY) have recently presented a new model for chemical evolution in the early
Galactic halo, explicitly taking into account the expected inhomogeneous nature
of the gas, and showed that their model is useful in interpreting the stellar
abundances of heavy elements originating from SNe II.

The light elements such as Be and B (BeB) in halo stars, which are the focus of
this paper, have usually been considered to originate from energetic protons
and $\alpha$-particles in Galactic cosmic rays (GCRs) colliding with heavy
nuclei of C, N, and O (CNO) in the interstellar medium (ISM) (e.g., Walker et
al. 1985).  While this secondary process naturally predicts a quadratic
dependence of BeB abundance on metallicity, newly obtained data of stellar BeB
abundances have, to the contrary, established a linear dependence (Boesgaard et
al. 1999 for Be; Duncan et al. 1997 and Primas et al. 1999 for B).

To account for this discrepancy, many authors have considered possible primary
processes for producing the BeB elements (e.g., Olive et al. 1994; Cass\'{e}
et al. 1995; Vangioni-Flam et al. 1996; Lingenfelter et al. 1998;
Vangioni-Flam et al. 1998; Higdon et al. 1998).  In particular, Duncan et
al. (1992) suggested that GCR CNO collide with protons and
$\alpha$-particles in 
the ISM to produce BeB, which leads to their linear dependence on metallicity
(see also Yoshii, Kajino, \& Ryan 1997; Ramaty et al. 1997). From a reanalysis
of the elemental abundances of present-day GCRs and an examination of the
physics of the SNR shock wave acceleration process, Meyer \&
Ellison (1999, hereafter ME99) and
Ellison \& Meyer (1999) have recently argued that most GCRs were accelerated
out of the ISM or circumstellar material instead of SN ejecta, although they
suggested that, owing to the lack of ISM CNO at early epochs, GCR CNO out of
fresh SN ejecta might play a role in producing the BeB elements.

In this paper, by incorporating both primary and secondary processes of GCR
spallation of heavy elements, we can, for the first time, construct a model for
the evolution of light elements which is fully consistent with the
inhomogeneous chemical evolution of the early Galactic halo.  The novelty of
the result based on this model is that the frequency distribution of metal-poor
stars in the log(BeB/H)--[Fe/H] plane is predictable, along with the
expected intrinsic abundance scatter.

\section{FORMULATION}
\label{sc:frml}

\subsection{Inhomogeneous Chemical Evolution}

We assume that the star-forming process is confined in separate clouds, 
which make up the entire halo of the early galaxy, and that the GCRs arise from
multiple SN explosions therein.  In our model a certain fraction $x_{\rm III}$
of cloud mass has turned into Pop III stars at time $t=0$ and the massive stars
among them explode as Pop III SNe to initiate the chemical evolution in a
given cloud.  All stars of subsequent generations are assumed to form from the
SNR shells behind the radiative shock front, which is supported by recent
observations of the abundance patterns in metal-poor stars (ST98).  The mass
fraction of each shell turning into stars is taken as a constant, denoted by
$\epsilon$.  Following TSY, provided $x_{\rm III}<1\times 10^{-2}$, a value of
$\epsilon=4.3\times 10^{-3}$ is used in the star formation rate given by
\begin{equation}
\label{eqn:sfr}
\hspace{-0.03cm}\displaystyle{
\dot{M_\ast}(t)}=\int_{\max(m_t{\rm ,}\,m_{{\rm SN},\,l})}^{m_u}
\hspace{-1.8cm}dm{\epsilon}M_{\rm sh}(m,t)\frac{\phi(m)}{m}
\dot{M_\ast}(t-{\tau}(m)){\rm ,}
\end{equation}
where ${\tau}(m)$ denotes the lifetime of a star with mass $m$, and $m_t$ 
is the stellar mass for which ${\tau}(m)=t$.  The initial mass function,
$\phi(m)$, having a Salpeter slope index of $-1.35$, is adopted with upper and 
lower mass limits of $m_u=50\Msun$ and $m_l=0.05\Msun$.  A lower mass limit 
of stars that explode as SNe is taken to be $m_{{\rm SN},\,l}=10\Msun$.  The
mass of the shell is given by $M_{\rm sh}(m,t)=M_{\rm ej}(m)+M_{\rm
sw}(m,t){\rm ,}$ where $M_{\rm ej}(m)$ is the mass of the SN ejecta and $M_{\rm
sw}(m,t)$ is the mass of the gas swept up by the SNR.  Since $M_{\rm
sw}(m,t)$ is insensitive to $m$ and
$t$, we here adopt $M_{\rm sw}(m,t)=6.5{\times}10^4{\Msun}$ (ST98; TSY).

The mass of the gas, $M_{\rm g}(t)$, and the abundance of the $j$-th heavy
element in the gas, $Z_{j,{\rm g}}(t)$, change with time according to eqs.(2)
and (3) in TSY, respectively.  Because of the inhomogeneous mixing of chemical
compositions in the dense shells and the ISM, the abundance of the $j$-th heavy
element in stars born at time $t$ in an SNR shell depends on the mass $m$ of SN
progenitor, and is essentially given by $Z_{j,\ast}(m,t)=[M_{Z_j}(m)+Z_{j,{\rm
g}}(t)M_{\rm sw}(m,t)]/M_{\rm sh}(m,t)$ (see TSY).

Since the light elements (LiBeB) are easily destroyed by stellar nuclear
reactions at temperatures above a few $10^6$ K, the abundance of the $L$
element in the gas $Z_{L,{\rm g}}(t)$ changes through destruction by stellar
lock-up and production by GCR reactions according to:
\begin{displaymath}
\hspace{-1cm}\displaystyle
{\frac{d(Z_{L, {\rm g}}M_{\rm g})}{dt}}=-\int_{\max 
(m_t,m_{{\rm SN}, l})}^{m_u}
\hspace{-1.8cm}dmZ_{L, \ast}(m,t){\epsilon}M_{\rm sh}(m,t)
\frac{\phi{(m)}}{m}
\end{displaymath}
\begin{displaymath}
\times\dot{M_\ast}(t-{\tau}(m))
+\sum_{i=p\alpha,j={\rm CNO}}\{\langle\sigma_{ij}^L
F_i\rangle Z_{j, {\rm g}}(t)(A_L/A_j)
\end{displaymath}
\begin{equation}
\label{eqn:chevle}
+\langle\sigma_{ji}^LF_j\rangle X_i(t)
(A_L/A_i)\}{M_{\rm g}(t)}{\rm ,}
\end{equation}
where $A_i$ is the mass number of the $i$-th element and $X_i$ is the 
abundance of hydrogen or helium.  We define 
$\langle\sigma_{ij}^LF_i\rangle\equiv\int_{E_{\rm th}}^{\infty}
\sigma_{ij}^L(E)F_i(E, t)S_L(E)dE$, where $\sigma_{ij}^L(E)$ is the 
cross section for the process of GCR projectile $i$ impinging on the 
ISM target $j$ to produce the $L$ element.  $S_L(E)$ gives the 
retention fraction of $L$ products that can survive to be thermalized 
into the ISM, and $F_{i}(E, t)$ is the time-dependent flux of GCR 
projectile $i$, which is modeled in \S \ref{ss:gcrfl}.  
The stellar $L$ abundance is equal to that in the shell and is given by  
\begin{displaymath}
\hspace{-1cm}\displaystyle
Z_{L,\ast}(m,t)=[Z_{L,{\rm g}}(t)M_{\rm sw}(m,t)+\sum_{i=p\alpha,j={\rm
CNO}}
\end{displaymath}
\begin{displaymath}
\{\langle\sigma_{ij}^LF_i\rangle Z_{j,{\rm g}}(t)
(A_L/A_j)
+\langle\sigma_{ji}^LF_j\rangle X_i(t)
(A_L/A_i)\}
\end{displaymath}
\begin{equation}
\label{eqn:shable}
\times{M_{\rm sh}(m,t)}{\Delta}T]/M_{\rm sh}(m,t) \;\;,
\end{equation}
where $\Delta T=10^6\: {\rm yr}$ is a typical diffusion time of a SNR shell
(Shigeyama 1999, private communication).  The first term in the numerator 
represents the mass of $L$ element swept up in the shell and the second 
term represents that mass produced during $\Delta T$ by GCRs originating 
from all the SNe that explode at time $t$.

\subsection{A Model of Time-Dependent GCR Flux}
\label{ss:gcrfl}
The energy spectrum of the flux of propagated GCR projectiles, denoted by
$F_i(E,t)$ in units of ${\rm particles\: s^{-1}cm^{-2}}$${\rm  (MeV\: per\:
nucleon)^{-1}}$, is expresssed as $F_i(E,t)=[\omega_i(E)]^{-1}\int_{E_{\rm
th}}^{\infty}q_i(E^{\prime},t)\exp[-(R_i(E^{\prime})-R_i(E))/\Lambda]dE
^{\prime}$ (Meneguzzi et al. 1971; Prantzos et al. 1993), where $q_{i}(E,t)$ is
the source spectrum in ${\rm particles{\:} s^{-1}g^{-1} (MeV{\:} per{\:}
nucleon)^{-1}}$, $\Lambda$ is the effective escape length in ${\rm g\:
cm^{-2}}$, ${\omega}_i(E)$ is the ionization energy losses in ${\rm MeV (g\:
cm^{-2}) ^{-1}}$ through a hydrogen-helium plasma with X=0.75 and Y=0.25, and
$R_{\rm i}(E)$ is the ionization loss length in ${\rm g\: cm^{-2}}$
(Northcliffe \& Schilling 1970).  The effective escape length is given by
$\Lambda^{-1}=\Lambda_{\rm e}^{-1}+\Lambda_{\rm n}^{-1}$ where $\Lambda_{\rm
e}$ is the loss length of energetic particles against escape from the Galactic
halo and $\Lambda_{\rm n}$ is that against nuclear destruction and decay.
${\Lambda}_e$ is proportional to the scale height $H$ in a cylindrical collapse
of the halo (Prantzos et al. 1993) and approximately changes with time as
$\Lambda_{\rm e}\propto 1-(at)^2$, where $a^{-1}$ is the timescale of the
collapse and a value of 1 Gyr is used below.  Although ${\Lambda}_e$ would be
huge in the past, the maximum value of $\Lambda$ is bounded by $\Lambda_{\rm
n}\simeq 200$ ${\rm g\: cm^{-2}}$ for protons and 20 ${\rm g\: cm^{-2}}$ for
$\alpha$ particles (Malaney \& Butler 1993).

We propose a new scenario in which GCRs originate from a combination of
injected SN ejecta (primary origin) and swept-up ISM (secondary origin) in the
SNR shell, which are then accelerated through the scattering back and forth
across the shock.  Their source spectrum $q_{i}(E,t)$ is assumed to have the
same form, with its intensity proportional to the ejection rate of energetic
particles associated with SNe, that is:
\begin{displaymath}
\hspace{-1cm}\displaystyle
q_i(E,t){\propto}(E+E_0)^{-2.6}\int_{\max ({m_t{\rm ,}\,
m_{SN,l}})}^{m_{u}}\hspace{-1.2cm}dm 
\end{displaymath}
\begin{equation}
\label{eqn:crfx}
\times\{M_{Z_i}(m)+Z_{i,{\rm g}}(t)f_{cr}M_{\rm sw}(m,t)\}
{\phi(m) \over A_im}{\dot M}_\ast ({t-\tau (m)}),
\end{equation} 
where $E_0=931{\rm MeV}$, $M_{Z_i}(m)$ is the mass of the $i$-th heavy 
element synthesized and ejected from a star with mass $m$, and $f_{cr}$ 
is the fractional mass of the gas in the shell swept up while the SN 
explosion is able to accelerate ISM particles.  If the acceleration were 
effective during the Sedov phase of the adiabatic shell expansion, a 
typical value of $f_{cr}$ would be $0.01-0.1$ (Taylor 1950; Sedov 1959).  
The absolute value of $q_i$ is chosen so that the predicted Be abundance 
from $F_i$ should agree with the observed abundance at [Fe/H]$=-3$.
The composition of $\alpha$ particles relative to protons in GCRs is
assumed to remain equal to a solar value of $F_\alpha/F_p=0.135$ (Lund 
1989), but the ratio $F_{\rm CNO}/F_p$ is calculated by using 
eq.(\ref{eqn:crfx}).

\section{RESULTS AND DISCUSSION}
\label{sc:res}
We have calculated the evolution of light and heavy elements in an
inhomogeneous Galactic halo using the formulae of the preceding section.  The
values of the free parameters chosen in our model are: $x_{\rm III}=10^{-6}$,
$\epsilon=4.3\times 10^{-3}$, and ${f_{cr}}=0.01$.  The initial abundances of
heavy elements are set to zero, but those of BeB elements are taken as equal to
the predicted primordial abundances of log(Be/H)=$-17.9$ and log(B/H)=$-16.9$,
based on the standard Big Bang nucleosynthesis (SBBN) model by Thomas et
al. (1994).

Our results are shown in the color-coded frequency distribution of
long-lived stars with $m<1\Msun$ in the log(Be/H)-[Fe/H] plane (Fig.1) and 
in the log(B/H)-[Fe/H] plane (Fig.2).  Both sets of frequency
distributions have been convolved with a typical error $\sigma=0.15$ dex in
the abundance measurements in order to directly compare with the Keck data for
Be (Boesgaard et al. 1999) and the HST data for B (Duncan et al. 1997; Primas
et al. 1999).

\begin{figure}[h]
\epsfxsize=8cm
\epsfbox{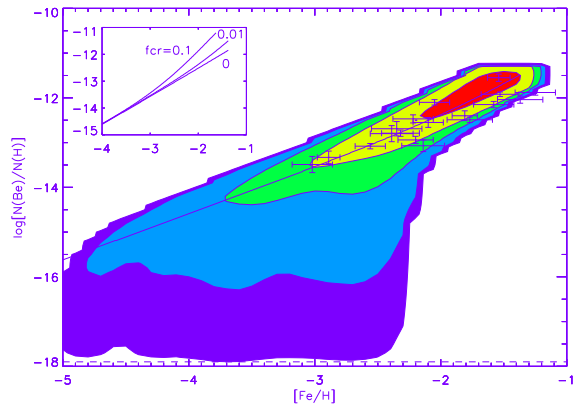}
\caption{Color-coded frequency distribution of long-lived 
stars in the log(Be/H)-[Fe/H] plane, convolved with a Gaussian having 
$\sigma=0.15$ dex for both log(Be/H) and [Fe/H].  The primorial Be 
abundance is taken from the SBBN model, and is indicated by the dashed line.  
The five contour lines, from the inside to the outside, correspond to those of
constant probability density $10^{-2}$, $10^{-3}$, $10^{-4}$, $10^{-6}$, and
$10^{-8}$ in unit area of $\Delta$[Fe/H]=0.1$\times\Delta$log(Be/H)=0.1.  The
thick line shows the log(Be/H)-[Fe/H] relation for the gas for our standard
choice of $f_{cr}=0.01$; the lines in the inset show those for various
values of $f_{cr}$.  The crosses represent the data with observational errors
taken from Boesgaard et al. (1999).}
\end{figure}

\begin{figure}[h]
\epsfxsize=8cm
\epsfbox{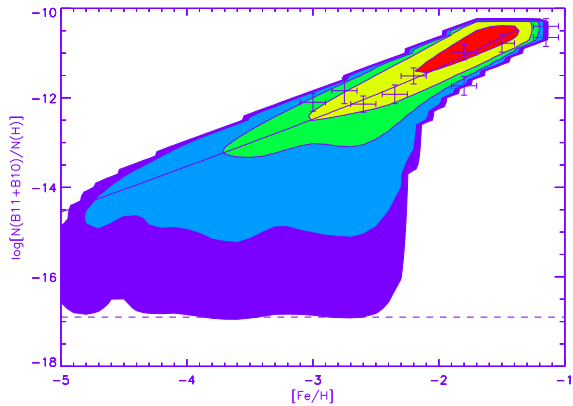}
\caption{Color-coded frequency distribution of long-lived 
stars in the log(B/H)-[Fe/H] plane, as in Fig.1 but for
B ($^{10}$B +$^{11}$B).  Crosses represent the data taken from Duncan et
al. (1997) and Primas et al. (1999)}
\end{figure}

We define a probability density of finding one halo star within a unit area of
$\Delta$[Fe/H]=0.1$\times\Delta$log(BeB/H) =0.1, normalized to unity when
integrated over the entire area.  The five contour lines shown in 
Figs.1 and 2 are, from the inside out, of constant probability density
$10^{-2}$, $10^{-3}$, $10^{-4}$, $10^{-6}$, and $10^{-8}$;  the total
probability integrated over the area inside each contour is 0.57, 0.95, 0.991,
0.9999, and 0.9999997, respectively.  Inspection of these figures shows that
the area of 0.95 probability is in reasonable agreement with the distribution
of observed data for both Be and B, and that, except for the presumed existence
of metal-free Pop III stars of the first generation, there should exist no
stars below [Fe/H]$\approx -5$ because the SNR shells at each star formation
site are enriched by the heavy-element ejecta from Pop III SNe.  We note, 
however, that this value on the lower bound in the stellar metallicity is
subject to the nucleosynthesis calculations adopted for SNe II.

The pattern of predicted abundance scatter in the log(BeB/H)-[Fe/H] plane
changes distinctly from lower to higher metallicity regions.  This transition
occurs at [Fe/H]$\sim -2.5$, below which the heavy-element mass of SN ejecta 
exceeds the heavy-element mass in the gas swept up in the SNR shell (ST98).  
Therefore, in the lower metallicity region ([Fe/H]$<-2.5$), stars formed at 
the same epoch from different SNR shells will exhibit a range of metallicities,
reflecting a different nucleosynthesis yield for each SN II with different
progenitor mass.  On the other hand, the GCRs, which are accelerated either
directly from each SN II or from swept-up ISM, can escape from the SNR shell at
high speed without producing BeB through spallative reactions in the
shell.  As a result, most of BeB are produced in the ambient ISM far outside
the shell, although the GCRs coming from outside impinge on targets in the
shell to produce a small amount of BeB there.  This accounts for the prediction
that stars with [Fe/H]$<-2.5$ show a large scatter in the heavy-element
abundance (along the horizontal axis), while exhibiting only a small scatter in
the BeB abundance (along the vertical axis).  It is therefore important to note
that purely GCR-induced products like BeB can still be used as an good age
indicator at early epochs, although the so-called age versus metallicity
relation is broken.

The predicted scatter in [Fe/H] becomes smaller towards higher [Fe/H], so
that the frequency distribution of stars in the log(BeB)-[Fe/H] plane 
converges to a linear relation BeB/H$\propto$Fe/H, as is observed over a range 
of $-3<$[Fe/H]$<-1$.  This linear trend emerges when the primary GCR 
processes produce BeB.  Notably, as suggested by ME99, the primary
processes by GCR CNO impinging on
ISM $p\alpha$, which are overwhelmed by the secondary processes at later
epochs (including the present), become dominant at early epochs because of the
lack of ISM CNO.  This does not mean that most GCRs are of the primary origin.
The mass ratio of primary to secondary GCRs in our model is $M_{\rm
ej}(m)/(f_{cr} M_{\rm sw}(m,t))\approx 10/(0.01\times 6.5\times
10^{4})\sim{0.02}$ (cf.
eq.(\ref{eqn:crfx})), which is not inconsistent with the recent results by
ME99 that the GCRs originate mostly from the ISM. This ratio could also
be argued by energetics between the kinetic energy released from SN and
the input energy onto GCRs(Ramaty et al. 1997), which will be discussed
in detail in our forthcoming paper.  

As [Fe/H] increases, the secondary GCR processes become dominant and a
quadratic relation BeB/H$\propto$ (Fe/H)$^2$ results.  This changeover from a
predicted linear to quadratic trend occurs at lower metallicity for larger
$f_{cr}$, as shown in the inset of Fig.1.  If a larger $f_{cr}$ is used,
in a typical range of $0.01-0.1$ (Taylor 1950; Sedov 1959), the predicted
relation deviates upwards from the observed linear trend.  This deviation is
partly offset by the effect of gas outflow from the cloud, which is known to
flatten the slope of the log(BeB/H)-[Fe/H] relation in the late phase of
chemical evolution (e.g., Yoshii et al. 1997).  In this paper we have taken
into account this outflow only approximately by using a lower value of
$f_{cr}=0.01$ as our standard.  More BeB data for stars in the
intermediate metallicity region $-2<$[Fe/H]$<-1$ are highly desirable to
constrain $f_{cr}$, that is, the acceleration mechanism of GCRs as well as the
rate of gas outflow from the cloud.

So far we have used the standard BBN model to set our assumed initial
abundances of BeB elements, although their abundances are strongly dependent on
the BBN model which is chosen.  For example, if the density distribution in the
early universe is not uniform as in the inhomogeneous BBN (IBBN) model, more
LiBeB are synthesized in neutron-rich regions (Kajino \& Boyd 1990).  In order
to see how a non-standard BBN model changes the result, we calculate the
frequency distribution of long-lived stars in the log(Be/H)-[Fe/H] plane
using an
optimal initial abundance of log(Be/H)=$-14.3$ taken from the IBBN model (Orito
et al. 1997).  The resulting distribution for the IBBN model is compared with
the SBBN model in Fig.3, where the contour lines of constant probability
density $10^{-4}$ and $10^{-5}$ in unit area of
$\Delta$[Fe/H]=0.1$\times\Delta$log(Be/H)=0.1 are shown.  Since the total
probability integrated over the area between the inner and outer contours is
less than 0.01, we need to measure BeB abundances in several hundred halo stars
in order to distinguish the IBBN from the SBBN on a statistical basis.  A more
direct way to distiguish these alternative hypotheses is to identify true Pop
III stars, the atmospheres of which consists of primeval gas.  According to
TSY, at least one halo star out of $10^{3-4}$ could be such a metal-free Pop
III star.  Such a large sample already exists (e.g., Beers 1999) and the
discovery of Pop III stars may not be an unrealistic wish in the near future.

\begin{figure}[h]
\epsfxsize=8cm
\epsfbox{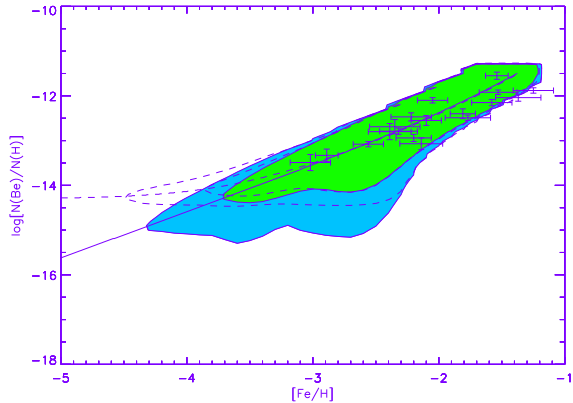}
\caption{Frequency distribution of long-lived stars in the log(Be/H)-[Fe/H]
plane for two different primordial Be abundances predicted in the SBBN and IBBN
models.  The contour lines, from the inside to the outside, correspond to those
of constant probability density $10^{-4}$ and $10^{-5}$ in unit area of
$\Delta$[Fe/H]=0.1$\times\Delta$log(Be/H)=0.1 for the SBBN (outer boundaries of
two regions painted by different colors) and for the IBBN (dashed lines).  The
log(Be/H)-[Fe/H] relation for the gas is shown by the thick line for the SBBN
and by the dashed line for the IBBN.  The crosses represent the data with
observational errors taken from Boesgaard et al. (1999).}
\end{figure}

\section{SUMMARY}

We have presented a general approach for studying the GCR-induced evolution of
the $L$ elements (LiBeB) in a manner consistent with an inhomogeneous chemical
evolution of the Galactic halo.  By adoption of a two-component source of GCRs,
our model results include: (1) a prediction of the intrinsic scatter in BeB and
[Fe/H] abundances within the model, (2) a successful prediction of the observed
linear trend between BeB and Fe, (3) a proposal for using BeB as a cosmic
clock, as an alternative to [Fe/H], and (4) a method for possibly constraining
the BBN model from future observations of metal-poor stars.

This work has been supported in part by the Grant-in-Aid for Scientific
Research (10640236, 10044103) and the COE research (07CE2002) of the 
Ministry of Education, Science, Sports and Culture of Japan.
We would like to thank T. C. Beers, T. Tsujimoto, and T. Shigeyama for
many fruitful discussions. \\

\end{document}